\documentclass[]{spie}  

\usepackage{booktabs}
\usepackage[final]{microtype}
\usepackage{amsmath,amsfonts,amssymb}
\usepackage{multirow}
\usepackage{graphicx}
\usepackage{subcaption}
\usepackage[colorlinks=true, allcolors=blue]{hyperref}

\title{AniRes2D: Anisotropic Residual-enhanced Diffusion\\ for 2D MR Super-Resolution}

\author[a]{Zejun~Wu}
\author[b]{Samuel~W.~Remedios}
\author[c]{Blake~E.~Dewey}
\author[a]{\\Aaron~Carass}
\author[a,b]{Jerry~L.~Prince}
\affil[a]{Department of Electrical and Computer Engineering, Johns~Hopkins~University,~Baltimore,~MD~21218,~USA}
\affil[ ]{\vspace*{-.75em}}
\affil[b]{Department of Computer Science, Johns Hopkins University, Baltimore,~MD~21218,~USA}
\affil[ ]{\vspace*{-.75em}}
\affil[c]{Department of Neurology, Johns Hopkins School of Medicine, Baltimore,~MD~21287,~USA}

\authorinfo{Further author information: (Send correspondence to Z.W.) Z.W.: E-mail: zwu76@jhu.edu}

\begin{document} 
\maketitle

\begin{abstract}

Anisotropic low-resolution~(LR) magnetic resonance~(MR) images are fast to obtain but hinder automated processing.
We propose to use denoising diffusion probabilistic models~(DDPMs) to super-resolve these 2D-acquired LR MR slices.
This paper introduces AniRes2D, a novel approach combining DDPM with a residual prediction for 2D super-resolution~(SR). 
Results demonstrate that AniRes2D outperforms several other DDPM-based models in quantitative metrics, visual quality, and out-of-domain evaluation.
We use a trained AniRes2D to super-resolve 3D volumes slice by slice, where comparative quantitative results and reduced skull aliasing are achieved compared to a recent state-of-the-art self-supervised 3D super-resolution method.
Furthermore, we explored the use of noise conditioning augmentation~(NCA) as an alternative augmentation technique for DDPM-based SR models, but it was found to reduce performance.
Our findings contribute valuable insights to the application of DDPMs for SR of anisotropic MR images.
\end{abstract}


\section{INTRODUCTION}
\label{sec:intro}
This paper focuses on the super-resolution~(SR) of anisotropic low-resolution~(LR) magnetic resonance~(MR) images~(MRIs).
Anisotropic LR scans are routinely acquired due to the impracticality of high-resolution~(HR) scans, as HR scans can suffer due to patient movement, are expensive and time consuming, and can also be limited by available hardware~\cite{du2020super}.
SR methods aim to address this issue by enhancing the quality of LR images, providing clearer views and potentially reducing the need for repeated scans~\cite{remedios2021joint, zhao2022medical, remedios2022deep, remedios2023self}. Importantly, SMORE4\cite{remedios2023self} is a recent self-supervised state-of-the-art method in anisotropic LR super-resolution, effectively addressing MR images with or without slice gaps.

Generative adversarial networks~(GANs)~\cite{goodfellow2014generative} have been used for SR in medical images~\cite{mahapatra2019image}, but their training is complex and prone to mode collapse~\cite{arjovsky2017wasserstein, ravuri2019classification}.
Denoising diffusion probablistic models~(DDPMs)~\cite{ho2020denoising, sohl2015deep} do not exhibit such issues, potentially enhancing the quality of super-resolved medical images~\cite{mao2023disc,kazerouni2023diffusion,wu2023super}, but there has been limited research on the performance of DDPMs for SR of anisotropic MR images. Additionally, faster sampling methods such as denoising diffusion implicit models~(DDIMs)~\cite{song2020denoising}, greatly
reduce the time needed during inference.

Recent advancements have been made in natural image super-resolution with DDPMs.
These include SR3~\cite{saharia2022image}, which employs DDPMs contitioned on LR images, to perform SR through a stochastic iterative denoising process. 
SRDiff~\cite{li2022srdiff} applies a residual prediction using DDPMs for faster convergence and greater stability during training.
SRDiff also interpolates a pretrained encoder on natural images to process the low resolution images before making a conditional noise prediction.
In this paper, we explore the effectiveness of integrating residual prediction into the process of super-resolving MR images, highlighting its impact on enhancing image quality and resolution.
Other work in the natural image domain include SR3\,+\,~\cite{sahak2023denoising} whish was designed for blind natural image SR, where multiple types of degradation and noise conditioning augmentation~(NCA) are applied.
In this paper, although various types of degradation are unsuitable for MR images, we explore NCA, which has the potential to mitigate domain shift caused by MR images from different sites and with different contrasts.
This potentially offers an interesting alternative to the issue of domain adaptation in MR images over the currently explored option of MR harmonization~\cite{zuo2023cmig}.

The performance of SR3, SR3 enhanced by residual prediction (referred to as AniRes2D), SR3 augmented by NCA (referred to as AniNCA2D) and SR3 with both residual prediction and NCA (referred to as ResNCA2D) in super-resolving anisotropic MR images are evaluated in this paper.
We apply a DDIM sampler to allow for fast sampling that meets practical needs.
Our experimental results comparing these four methods show the superiority of AniRes2D quantitatively and qualitatively.
Additionally, AniRes2D exhibits the best performance on out-of-domain evaluation. 
Unexpectedly, NCA~\cite{sahak2023denoising} degrades both in-domain and out-of-domain performance, as detailed in Section~\ref{sec:experiments}.
AniRes2D is also applied to both sagittal and coronal slices.
By stacking and averaging these enhanced slices, our results demonstrate quantitative performance comparable to the state-of-the-art self-supervised 3D super-resolution method, SMORE4~\cite{remedios2023self}, with enhanced reconstruction in
Fourier space and reduced aliasing effects on the skull, showing the promise of the DDPM-based methods on anisotropic MR image super resolution.

\section{Methods}
\label{sec:method}
\textbf{Problem Formulation}\quad Given a dataset \( D = \{ (x_i, y_i)\}_{i=1}^{N} \), where \( x_i \) are anisotropic LR images and \( y_i \) are HR images from an unknown conditional distribution \( p(y | x) \), the goal is to approximate this distribution via a mapping \( f \) from LR to HR:
\begin{align*}
\text{Find } f: \mathbb{R}^{m \times n} \rightarrow \mathbb{R}^{km \times n} \text{ such that } f(x_i) \approx y_i, \forall i = 1, \ldots, N \text{ assuming } (x_i, y_i) \sim p(y|x)\,,
\end{align*}
where $k$ is the SR factor between the LR and HR images.
To approximate $f$, four DDPM-based methods, SR3, AniRes2D, AniNCA2D, and ResNCA2D are used. DDIMs are used to speed up the sampling of our DDPM-based methods. Anisotropic LR MRI slices are characterized by a slice thickness of $t$~mm and a gap of $g$~mm (denoted as $t\|g$ following Remedios~\textit{et al.}~\cite{remedios2022deep}), which results in $k$ being equal to $(t+g)$. 
We simulate acquisition with the Shinnar–Le Roux pulse design~\cite{martin2020sigpy}.

\subsection{SR3}\quad SR3 approximates the unknown conditional distribution \( p(y | x) \) by an iterative refinement process, followed by a noise image drawn from a Gaussian distribution \( y_T \sim \mathcal{N}(0, I) \).
The image is then refined over \( T \) stages, resulting in the target image \( y_0 \).
The iterative process can be described by:
\begin{align*}
y_T \sim \mathcal{N}(0, I), y_{t-1} \sim p_\theta(y_{t-1} | y_t, x_i), \text{ for } t = T, \ldots, 1,
\end{align*}
where \( p_\theta(y_{t-1} | y_t, x_i) \) are learned conditional transition distributions, and \( y_0 \sim p(y | x_i) \).
The recovery of the signal from noise is achieved through a reverse Markov chain conditioned on the LR image \( x_i \), and a neural denoising model \( f_\theta \) is used to learn the reverse chain. In experiments, we condition on the upsampled version of LR image  \( U(x) \) instead of \( x_i \). Bicubic interpolation is used for upsampling, implemented using \texttt{scipy.ndimage.zoom} function from the SciPy library\cite{2020SciPy-NMeth}.

\subsection{AniRes2D, AniNCA2D and ResNCA2D}
\begin{figure}[!tb]
   \centering
   \includegraphics[height=5cm]{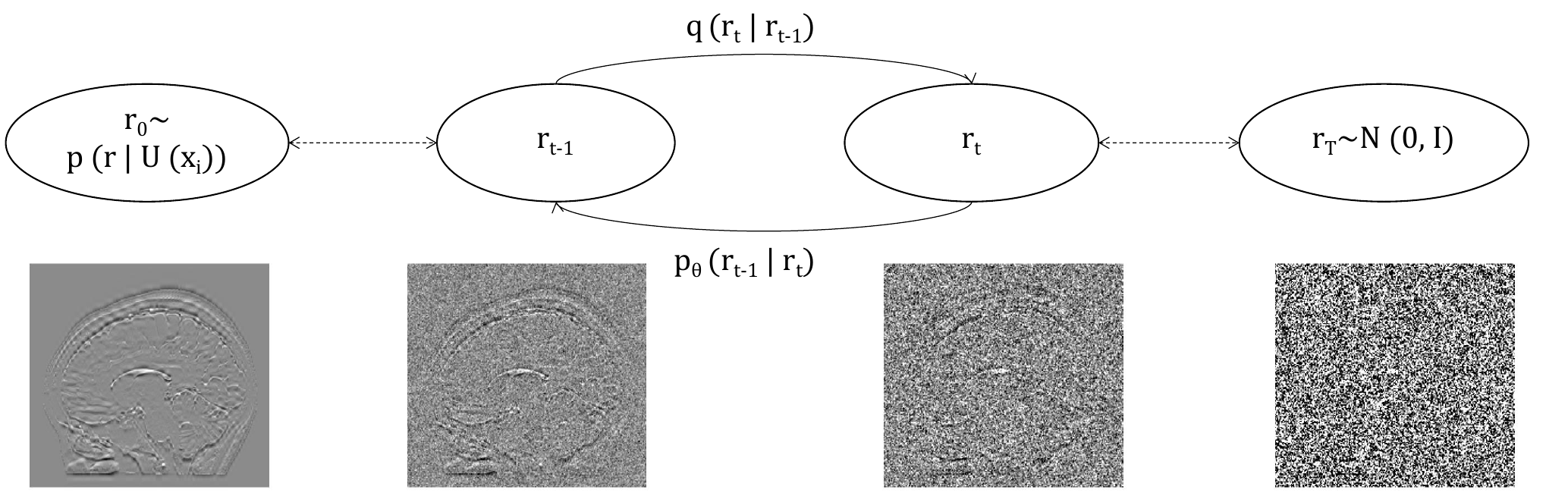}
   \caption[example]
   { \label{fig:anires2d} Illustration of AniRes2D. In the forward diffusion procedure, denoted by \( q \) and progressing from left to right, the process incrementally introduces Gaussian noise into the desired image. Conversely, the backward inference mechanism, represented by \( p \) and moving from right to left, systematically removes noise from the desired image, doing so in a manner that depends on an upsampled LR image \( U(x_{i}) \). The upsampled LR image \( U(x_{i}) \) is not depicted in this illustration.
}
\end{figure}

\textbf{AniRes2D}\quad AniRes2D applies a residual prediction to target the high-frequency details typically missing in LR images. Unlike SR3, which directly generates HR images, AniRes2D focuses on generating the residual space between the HR and upsampled LR images. The process begins with upsampling the LR image \( x \), and we get \( U(x) \).
The formulation can be expressed as:
\[
r_i = y_i - U(x_i), \text{ for } i = 1, \ldots, N; r_T \sim \mathcal{N}(0, I); r_{t-1} \sim p_\theta(r_{t-1} | r_t, U(x_i)), \text{ for } t = T, \ldots, 1; \]\[r_0 \sim p(r | U(x_i)); y_0 = r_0 + U(x_i).
\]
Here \( r_i \) is the residual between the true HR image \( y_i \) and the upsampled LR image.
By focusing on this residual prediction, AniRes2D forces the model to capture the high-frequency details, enhancing the quality of SR.
An illustration of the process is shown in Fig.~\ref{fig:anires2d}.

\textbf{AniNCA2D}\quad AniNCA2D integrates noise conditioning augmentation~(NCA) with SR3, which theoretically enhances the robustness of the SR model, especially enhancing the model's ability on unseen data.
The process of AniNCA2D encompasses the following steps:
(1)~Upsample the LR image: \( U(x_i) \);
(2)~Sample a noise level \(\tau \sim \text{Uniform}(0, \tau_{\text{max}} = 0.5)\), and create a noise sample \( c_\tau \sim \mathcal{N}(0, I) \);
(3)~Add noise to the up-sampled LR input: \( x_{\text{noisy}} = U(x_i) + c_\tau \); and 
(4)~Initiate the iterative refinement process with a noise image \( y_T \sim \mathcal{N}(0, I) \).
\[
  y_T \sim \mathcal{N}(0, I), \quad y_{t-1} \sim p_\theta(y_{t-1} | y_t, x_{\text{noisy}}), \quad \text{ for } t = T, \ldots, 1, \quad y_0 \sim p(y | x_{\text{noisy}}),
\]

\textbf{ResNCA2D}\quad ResNCA2D integrates SR3 withNCA, the residual prediction, and the upsampling procedure.
For the sake of brevity and to avoid repetition, a detailed description of this model has been omitted.

\section{Experiments}
\label{sec:experiments}

\subsection{Data}
MR volumes from the following datasets are used in our experiments. 
The \textbf{\textit{3D-MR-MS Dataset}}~\cite{lesjak2018novel}~comprises 3D FLAIR images from 30 people with multiple sclerosis~(PwMS) with observable white matter lesions.
The \textbf{\textit{OASIS3 Dataset}}~\cite{lamontagne2019oasis}~is a vital resource for aging and Alzheimer's research, providing a diverse range of brain MR images and longitudinal data that are essential for understanding age-related neurological changes.
The \textbf{\textit{IXI Dataset}}\footnote{https://brain-development.org/ixi-dataset/} is a collection of 600 MR brain images from normal, healthy subjects, which includes T1-weighted~(T$_2$-w) and  T2-weighted~(T$_2$-w) images.
The images were acquired at three different hospitals in London and are available under the Creative Commons CC BY-SA 3.0 license.
The \textbf{\textit{Human Connectome Project (HCP) Dataset}}~\cite{van2013wu}~provided by WU-Minn Consortium, includes data from the NIH Blueprint for Neuroscience Research and the McDonnell Center for Systems Neuroscience at Washington University. In our experiments, we focus on super-resolve $3\|1$ LR images, so $t$ is 3, $g$ is 1 and $k$ is 4 in all of our experiments.

\subsection{Evaluation on 2D slices}
\label{subseq:eval_2d}
In this subsection, we train on pairs of 2D HR and LR slices and evaluate only on 2D slices, since we focus on performance on 2D slices here.
From \textit{3D-MR-MS}, we used 26 PwMS into the training dataset, 1 PwMS in the validation dataset, and 3 PwMS in the inference dataset.
Sagittal and coronal slices pulled from axially-acquired LR, sagittal and axial slices pulled from coronally-acquired LR, and coronal and axial slices pulled from sagittally-acquired LR, together with their corresponding HR slices, are the HR-LR pairs in the experiments.
Additionally, $10$ T$_1$-w subjects were randomly selected from the \textit{OASIS3 Dataset} to test the domain adaptation ability of the model.
For \textit{OASIS3}, only $10$ center sagittal and coronal slices on axial LR images were extracted, yielding 200 pairs.

We applied linear normalization to $[0, 1]$ using the max and min values of the LR to both HR and LR images.
Reflect padding was employed to align the image size with the dimensional requirements of the U-net architecture.
Due to limited samples, affine transformations, elastic transformations, and flipping were used.
\begin{table}[!tb]
    \centering
    \caption{\textit{3D-MR-MS} results. The best result for each metric is marked in bold. \textbf{Key:} BI - bicubic interpolation; AR - AniRes2D; AN - AniNCA2D; RN - ResNCA2D; w/ NCA - with NCA during inference; w/o NCA - without NCA during inference.}
    \label{tab:3dmrms_results}
    \begin{tabular}{l *{7}{l}}
        \toprule
        \textbf{Metric} & \textbf{BI} & \textbf{SR3} & \textbf{AR} & \textbf{AN} & \textbf{AN} & \textbf{RN} & \textbf{RN} \\
        &&&& \textbf{w/ NCA} & \textbf{w/o NCA} & \textbf{w/ NCA} & \textbf{w/o NCA}\\
        \cmidrule(lr){1-8}
        PSNR & 25.2861 & 10.8076 & \textbf{26.0434} & 10.5459 & 10.5710 & 25.2781 & 25.4582 \\
        SSIM & 0.8482 & 0.0827 & \textbf{0.8644} & 0.0718 & 0.0743 & 0.8383 & 0.8445 \\
        \bottomrule
    \end{tabular}
\end{table}

\begin{figure}[!tb]
   \centering
   \includegraphics[height=7cm]{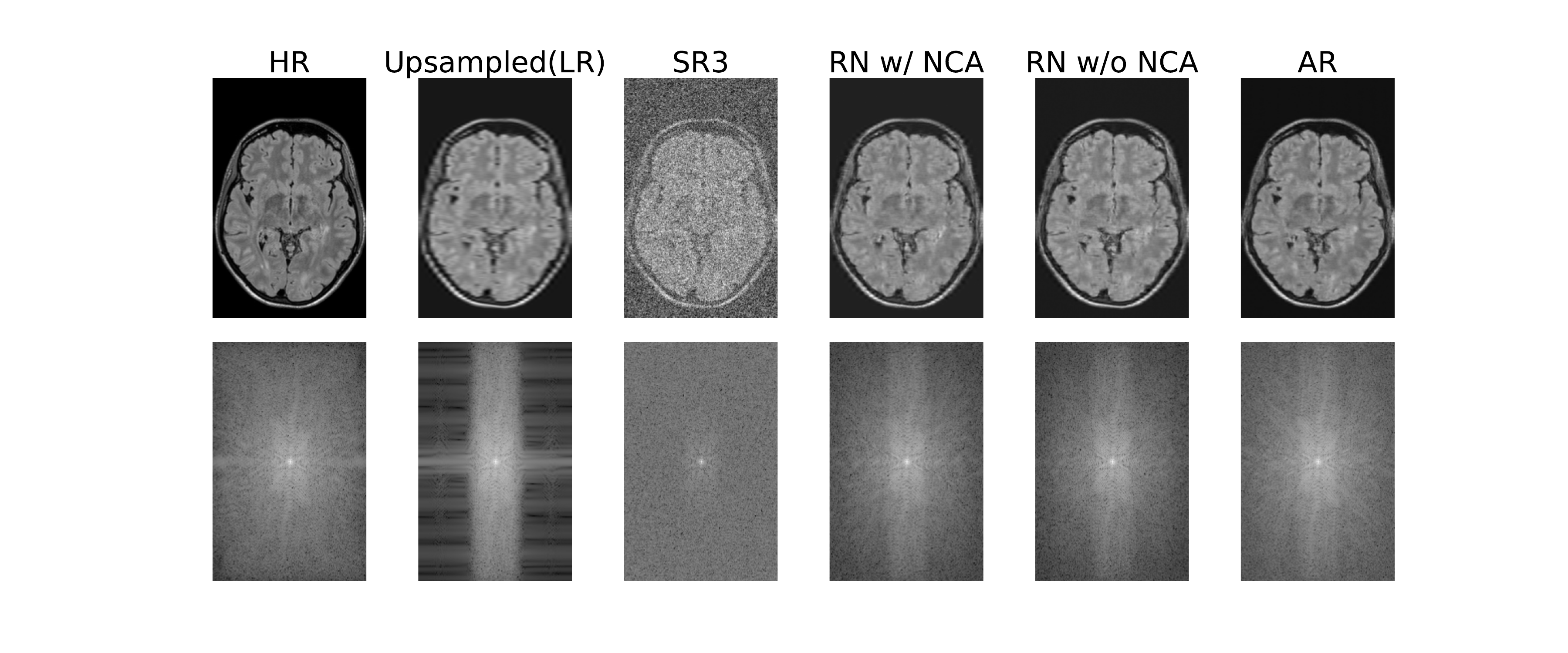}
   \caption[example]
   { \label{fig:example} 2D image slices and their corresponding Fourier space from the \textit{3D-MR-MS} dataset. Top row: HR, Upsampled LR, and super-resolved images (AniNCA2D omitted). Bottom row: corresponding Fourier spectra.}
\end{figure}

In Table~\ref{tab:3dmrms_results} we present in-plane evaluations including peak signal-to-noise ratio~(PSNR) and structural similarity index~(SSIM) for the \textit{3D-MR-MS} dataset.
The reported PSNR and SSIM represent the average of these metrics computed across all two-dimensional slices, comparing the SR images with their corresponding HR counterparts.
We show the magnitude of the Fourier transform in Fig.~\ref{fig:example}.
As seen in Table~\ref{tab:3dmrms_results} and Fig.~\ref{fig:example}, AniRes2D has the best results.
While it was initially expected that incorporating NCA during training would enhance performance, it surprisingly led to a degradation of results.
This observation aligns with findings from the original NCA paper~\cite{sahak2023denoising}, which reported better results when NCA was omitted during inference.
SR3 and AniNCA2D fail to estimate the HR images and show poor results, generating noisy SR images (AniNCA2D omitted for brevity in Figs.~\ref{fig:example} as its results are similar to SR3), which underscores the need for a residual prediction.

\begin{table}[!tb]
    \centering
    \caption{\textit{OASIS3} results. The best result for each metric is marked in bold. See Table~\ref{tab:3dmrms_results} for the key.}
    \label{tab:oasis3_results}
    \begin{tabular}{l *{7}{l}}
        \toprule
        \textbf{Metric} & \textbf{BI} & \textbf{SR3} & \textbf{AR} & \textbf{AN} & \textbf{AN} & \textbf{RN} & \textbf{RN} \\
        &&&& \textbf{w/ NCA} & \textbf{w/o NCA} & \textbf{w/ NCA} & \textbf{w/o NCA}\\
        \cmidrule(lr){1-8}
        PSNR & 25.6646 & 11.1202 & \textbf{25.7630} & 10.9041 & 10.9706 & 25.1331 & 25.2280 \\
        SSIM & \textbf{0.7998} & 0.1012 & 0.7923 & 0.0925 & 0.0983 & 0.7665 & 0.7700 \\
        \bottomrule
    \end{tabular}
\end{table}

We examined the model's domain adaptation performance using T$_1$-w slices from \textit{OASIS3} dataset.
Quantitative results are presented in Table~\ref{tab:oasis3_results}.
AniRes2D achieves the best PSNR and outperforms all other DDPM-based methods in terms of SSIM, showing its excellent domain adaptation ability.

NCA, which theoretically should enhance  robustness on unseen data, has failed to demonstrate empirical success in our setting.
We propose two potential reasons for this discrepancy.
Firstly, NCA was originally designed for natural images, characterized by an extensive array of textures, colors, and patterns, and often employs Gaussian noise.
Medical images, however, are often single channel as with MR and encapsulate specific structural information.
The introduction of unnecessary noise may disrupt these delicate structures and boundaries, complicating feature extraction.
Secondly, medical images often exhibit anisotropic properties, with differing resolutions across various directions. 
This non-uniformity may further complicate the application of Gaussian noise, making it detrimental to the image quality.
Consequently, we posit that the application of NCA to medical images necessitates careful consideration and tailoring.

\begin{figure}[!tb]
   \centering
   \includegraphics[height=4cm]{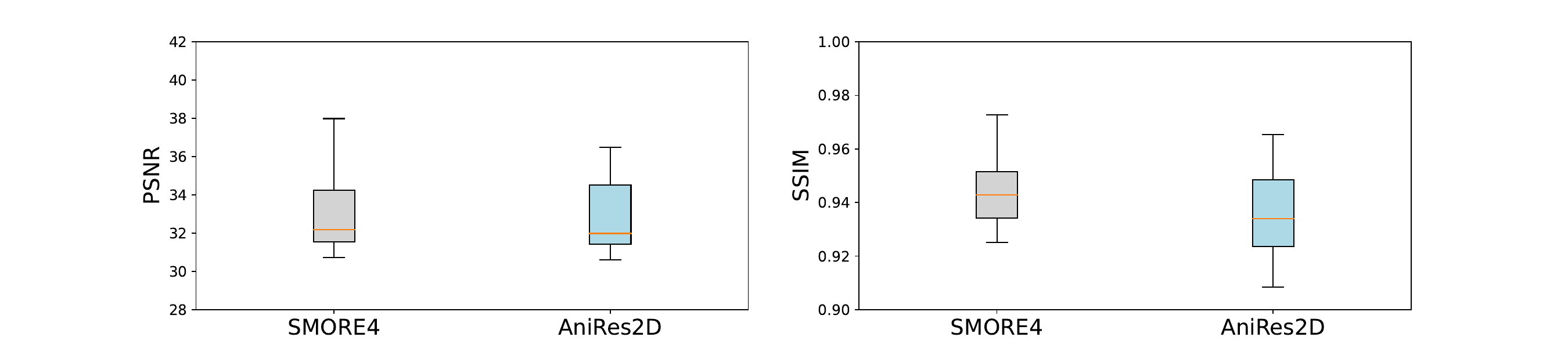}
   \caption[example]
   { \label{fig:3d_evaluation} Comparative performance analysis of SMORE4 and AniRes2D methods in terms of PSNR and SSIM metrics. The left subplot displays the performance in PSNR, with box plots representing SMORE4 (in light gray) and AniRes2D (in light blue). The right subplot shows the performance in SSIM, following the same color scheme.}
\end{figure}

\subsection{Comparison of Whole 3D Volume Evaluation with SMORE4}
We focus on the practical use of super-resolving 3D axially-LR volumes here. 
We trained two separate instances of AniRes2D, one for axial and the other for sagittal slices.
The final 3D volume was generated by stacking and averaging the super-resolved axial and sagittal outputs, aligning with the strategy used in Remedios \textit{et al.}~\cite{remedios2023self}.
We use a hybrid dataset consisting of 99 T$_1$-w or T$_2$-w volumes sourced from the \textit{OASIS3}, \textit{IXI}, and \textit{HCP} datasets, as shown in Table~\ref{tab:data_split}.

\begin{figure}[!tb]
   \centering
   \includegraphics[height=7cm]{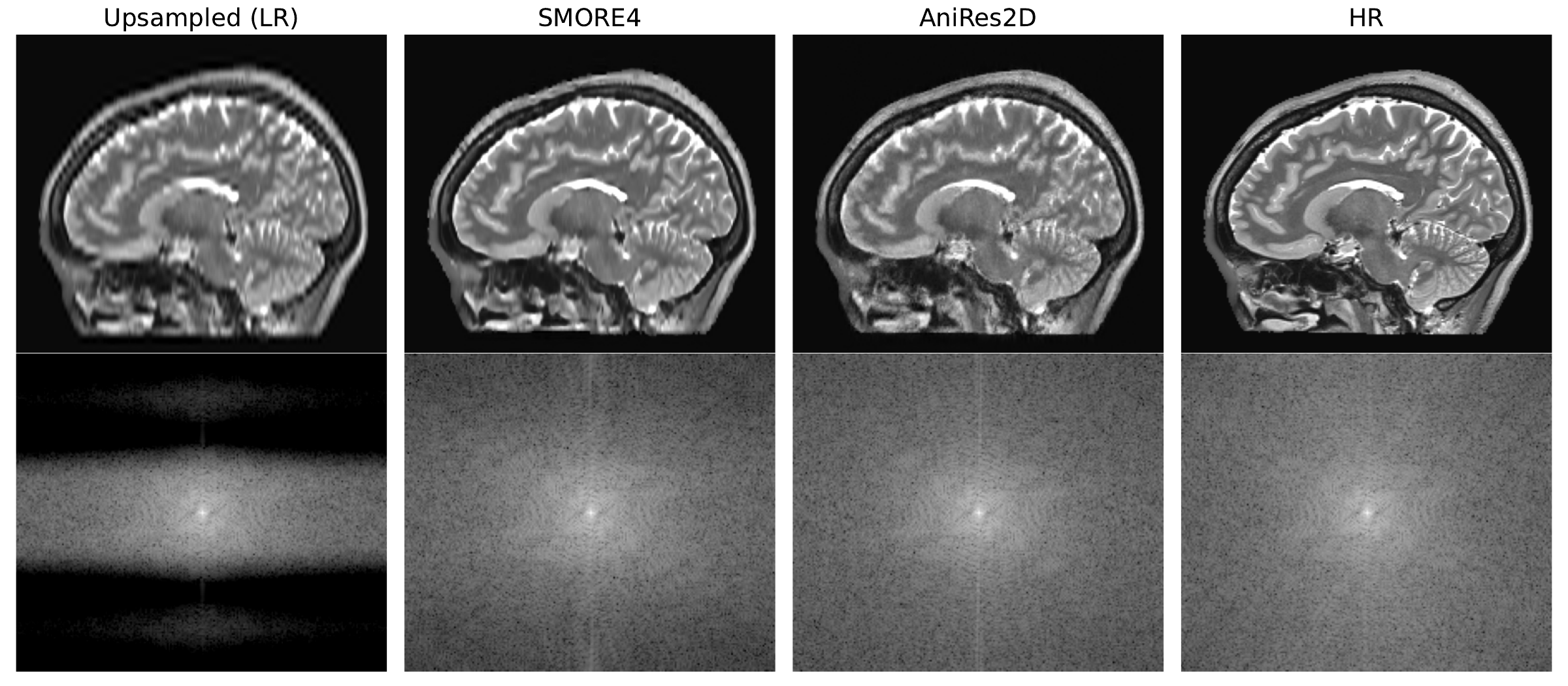}
   \caption[example]
   { \label{fig:sagittal} Sagittal super-resolved results for axially-acquired LR. Top row: Upsampled LR, super-resolved images and HR. Bottom row: corresponding Fourier spectra.}
\end{figure}

We use the same normalization  process as in Section~\ref{subseq:eval_2d}.
All sagittal and coronal 2D slices are extracted from axially-acquired LR volumes.
Reflect-padding is used to fit images to U-net networks.
Only flipping was used since it is a relatively large dataset. Results are shown in Figs.~\ref{fig:3d_evaluation} and~\ref{fig:sagittal}.

\begin{table}[!tb]
\centering
\caption{Distribution of MRI volumes used for training, validation, and testing from the OASIS3, HCP, and IXI datasets.}
\label{tab:data_split}
\begin{tabular}{l c cc c cc c cc}
\toprule
\multirow{2}{*}{\textbf{Dataset}} && \multicolumn{2}{c}{\textbf{Train}} && \multicolumn{2}{c}{\textbf{Validate}} && \multicolumn{2}{c}{\textbf{Test}} \\ 
\cmidrule(lr){3-4}
\cmidrule(lr){6-7}
\cmidrule(lr){9-10}
&\hspace*{2em}& \textbf{T$_1$-w} & \textbf{T$_2$-w} &\hspace*{1em}& \textbf{T$_1$-w} & \textbf{T$_2$-w} &\hspace*{1em}& \textbf{T1w} & \textbf{T$_2$-w}\\
\cmidrule(lr){1-10}
\textbf{OASIS3} && 36 & 0 && 4 & 0 && 9 & 0\\
\textbf{HCP}    && 7  & 7 && 1 & 1 && 2 & 2\\
\textbf{IXI}    && 21 & 0 && 3 & 0 && 6 & 0\\
\cmidrule(lr){1-10}
%
%
\textbf{Total}  && 64 & 7 && 8 & 1 && 17 & 2\\
\bottomrule
\end{tabular}
\end{table}

As can be observed in Figure \ref{fig:3d_evaluation}, for PSNR, the values for SMORE4 and AniRes2D range between approximately 30 to 38, and 31 to 36 respectively.
The average volumetric PSNR is 32.9020 for SMORE4 and 32.7272 for AniRes2D.
Both methods demonstrate high image reconstruction quality in PSNR terms, but SMORE4 shows slightly higher peaks and a broader range, suggesting potentially superior reconstruction in certain scenarios.
For SSIM: SMORE4 and AniRes2D mostly hover around 0.9 to 0.97.
The average volumetric SSIM is 0.9440 for SMORE4 and 0.9354 for AniRes2D.
Both methods achieve high SSIM values, indicating effective structural information preservation.
While SMORE4 reaches marginally higher SSIM in some instances, the overall performance is similar between the two. 
In summary, both SMORE4 and AniRes2D exhibit high image reconstruction quality across PSNR and SSIM metrics. 
Although SMORE4 slightly outperforms in some aspects, the performance difference is not significant. 

In terms of time efficiency, our comparison is conducted on an NVIDIA Quadro RTX 8000 with 48 GB of GPU memory. For SMORE4, the training duration for a single volume is approximately 22 minutes, which constitutes the majority of its super-resolution processing time. In contrast, two separate AniRes2D models are trained for sagittal and coronal slices, each requiring roughly 3 days of training time. When employing a DDIM sampler for 
$ 10\times$ faster sampling, it takes about 42 minutes to super-resolve a single volume, which is the primary time-consuming step in its super-resolution process. Empirical evidence suggests that DDIMs can generate high-quality samples at a rate $10\times$ to $50\times$ faster in wall-clock time compared to DDPMs \cite{song2020denoising}. This efficiency enables a flexible trade-off between computational resources and sample quality when necessary.

A qualitative comparison is shown un Figs.~\ref{fig:sagittal}, AniRes2D effectively eliminates aliasing artifacts in the skull regions, achieving superior reconstruction in Fourier space. Apart from these improvements, the overall image quality is remarkably similar between the compared methods.

\section{Conclusion}
\label{sec:conclusion}
In this work, DDPMs are used to super-resolve LR anisotropic MR images. DDIM samplers are adopted to allow for acceptable sampling time. We show that our strategy for whole 3D volume super-resolution achieves comparable quantitative results compared to a recent state-of-the-art self-supervised method, SMORE4. In our comparison, our strategy excels at reducing skull aliasing effects and reconstructing better Fourier Space.

Additionally, two strategies to enhance DDPMs are explored. Our empirical results, tested on both the original \textit{3D-MR-MS} dataset and the \textit{OASIS3} dataset, demonstrate that the integration of a residual prediction can substantially improve inference performance. We also investigated NCA, another contemporary technique for augmenting DDPMs in SR tasks. NCA is found to harm the image quality, and we listed potential reasons for that.

\bibliography{report} 
\bibliographystyle{spiebib} 

\end{document}